\documentclass{emulateapj}
\usepackage{apjfonts}
\usepackage{graphicx}
\usepackage{amsmath}

\begin{document}

\submitted{The Astrophysical Journal Letters, accepted}   
\vspace{1mm}




\title{Gamma Ray Flares from Mrk421 in 2008 observed with the ARGO-YBJ detector}


\author{G.~Aielli\altaffilmark{1,2},
 C.~Bacci\altaffilmark{3,4},
 B.~Bartoli\altaffilmark{5,6},
 P.~Bernardini\altaffilmark{7,8},
 X.J.~Bi\altaffilmark{9},
 C.~Bleve\altaffilmark{7,8},
 P.~Branchini\altaffilmark{4},
 A.~Budano\altaffilmark{4},
 S.~Bussino\altaffilmark{3,4},
 A.K.~Calabrese Melcarne\altaffilmark{10}
% ...
\\ (The Argo-YBJ Collaboration)}


 \altaffiltext{1}{Dipartimento di Fisica dell'Universit\`a ``Tor Vergata''
                  di Roma, via della
                  Ricerca Scientifica 1, 00133 Roma, Italy.}
 \altaffiltext{2}{Istituto Nazionale di Fisica Nucleare, Sezione di
                  Roma Tor Vergata, via della Ricerca Scientifica 1, 00133 Roma, Italy.}
 \altaffiltext{3}{Dipartimento di Fisica dell'Universit\`a ``Roma
                  Tre'' di Roma, via della Vasca Navale 84,
                  00146 Roma, Italy.}
 \altaffiltext{4}{Istituto Nazionale di Fisica Nucleare, Sezione di
                  Roma Tre, via della Vasca Navale 84, 00146 Roma, Italy.}
 \altaffiltext{5}{Dipartimento di Fisica dell'Universit\`a di Napoli,
                  Complesso Universitario di Monte Sant'Angelo,
                  via Cinthia, 80126 Napoli, Italy.}
 \altaffiltext{6}{Istituto Nazionale di Fisica Nucleare, Sezione di
                  Napoli, Complesso Universitario di Monte
                  Sant'Angelo, via Cinthia, 80126 Napoli, Italy.}
 \altaffiltext{7}{Dipartimento di Fisica dell'Universit\`a del Salento,
                  via Arnesano, 73100 Lecce, Italy.}
 \altaffiltext{8}{Istituto Nazionale di Fisica Nucleare, Sezione di
                  Lecce, via Arnesano, 73100 Lecce, Italy.}
 \altaffiltext{9}{Key Laboratory of Particle Astrophysics, Institute 
of High
                   Energy Physics, Chinese Academy of Science,
                   P.O. Box 918, 100049 Beijing, P.R. China.}
\altaffiltext{10}{Istituto Nazionale di Fisica Nucleare - CNAF, Viale 
                  Berti Pichat 6/2, 40127 Bologna, Italy.}
% ...


\begin{abstract}

In 2008 the blazar Markarian 421 entered a very active phase and was one
of the brightest sources in the sky at TeV energies, showing 
frequent flaring episodes. Using the data of ARGO-YBJ, 
a full coverage air shower detector located at
Yangbajing (4300 m a.s.l., Tibet, China), we monitored the source
at gamma ray energies E $>$ 0.3 TeV during the whole year. 
The observed flux was variable, with the strongest flares  
in March and June, in correlation with X-ray enhanced activity. 
While during specific episodes the TeV flux could be several times
larger than the Crab Nebula one, 
the average emission from day 41 to 180 was almost twice the Crab level, 
with an integral flux of (3.6$\pm0.6$) 
$\times$ 10$^{-11}$ photons cm$^{-2}$ s$^{-1}$ for energies E $>$ 1 TeV,
and decreased afterwards.

This paper concentrates on the flares occurred in the first half of June.
This period has been
deeply studied from optical to 100 MeV gamma rays, and partially
up to TeV energies, since the moonlight hampered 
the Cherenkov telescope observations during the most intense
part of the emission.
Our data complete these observations, with the detection of a signal
with a statistical significance of 3.8 standard deviations on June 11-13, 
corresponding to a gamma ray flux 
about 6 times larger than the Crab one above 1 TeV.

The reconstructed differential spectrum,
corrected for the intergalactic absorption, 
can be represented by a power law with an index 
$\alpha = -2.1^{+0.7}_{-0.5}$ extending up to several TeV.
The spectrum slope is fully consistent with previous observations 
reporting a 
correlation between the flux and the spectral index, 
suggesting that this property
is maintained in different epochs and characterizes the source emission
processes.

\end{abstract}


\keywords{BL Lacertae objects: individual (Markarian 421) - gamma rays: 
observations}  



\section{Introduction}

The blazar Markarian 421 has been the first extragalactic source
observed at gamma ray energy E$>$500 GeV \citep{Pun92}. 
It belongs to the radio-loud Active Galactic Nuclei (AGNs) 
sub class of BL Lacertae objects, characterized by
a non thermal spectrum extending up to high energies
and by a rapid flux variability at nearly all wavelenghts.

To date about 30 BL Lacs have
been detected at very high energies (VHE, E$>$100 GeV) and Mrk421
is the closest one (z = 0.031). 
Its relatively small distance
makes it one of the best studied TeV gamma ray sources.
Since its discovery this object played a significant role in the discussion
concerning both the emission processes in AGNs and the attenuation
of TeV gamma rays in the extragalactic space.

It is now widely recognised that the BL Lac radiation
originates in a relativistic jet pointing at a small angle to the line of sight
and that it is amplified by relativistic effects,
explaining both the strong high energy emission and its rapid variability.

Usually the BL Lacs energy density spectra have two broad band components, 
the first one peaking in the infrared to X-ray region,
the second one in the MeV-TeV range \citep{Sam96,Fos98}.
Mrk421 is classified as
a High-energy peaked BL Lac (HBL), showing
the peaks in the X-ray and VHE regions, respectively \citep{Pad95}.

The low energy component is commonly believed to 
originate as synchrotron emission
from relativistic electrons gyrating in the magnetic field of the jet plasma,
while the origin of the second one is still unclear.
Many models propose that gamma rays are produced in Inverse Compton 
scattering of synchrotron (Synchrotron Self-Compton, SSC) 
or ambient photons (external Compton, EC) off the same electron population
that produces the synchrotron radiation \citep{Ghi98,Der92}.
Alternatively, in the ``hadronic'' models, gamma rays are emitted 
as sychrotron radiation of extremely energetic protons, or by
secondary particles produced by protons interacting
with some target material \cite{Muc03}.

Flaring activity of Mrk421 at VHE
energies has been observed with variability time scales ranging
from minutes to months, and many multiwavelength campaigns have revealed 
a strong correlation of gamma rays with X-rays, 
that can be easily interpreted in terms of the SSC model \citep{Fos08,Wag08}.
In addition some data have shown significant variations 
of the TeV spectrum slope during different activity phases,
and an evident 
correlation between the spectral hardness and the flux intensity \citep{Kre02}.

The simultaneous observation at different wavelengths is of great
importance since may provide unique information about the source
properties and the radiation processes.

A set of measurements \citep{Don09} 
covering 12 decades of energy, from optical to TeV
gamma rays, was performed during the strong flaring activity in
the first half of June 2008 by different detectors: WEBT (optical R-band), UVOT
(UV), RXTE/ASM (soft X-rays), SWIFT (soft and hard X-rays), AGILE
(hard X-rays and gamma rays) and the Cherenkov telescopes VERITAS and
MAGIC (VHE gamma rays). These data allowed for a deep analysis of
the broad band energy spectrum as well as for the study of time
correlations among the fluxes in different energy ranges. 

In this period two flaring episodes were reported, the first one on June
4-8, observed from optical to TeV gamma rays, the second one,
larger and harder, on June 10-14, observed from optical to 100 MeV
gamma rays. Using the multifrequency data, Donnarumma et al. (2009)
derived the Spectral Energy Distribution (SED) 
for June 6, that shows the typical two humps
shape.
In the framework of the SSC model, according to the authors, 
the second hump intensity, that reached a flux of about 3 
$\times$ 10$^{-11}$ photons cm$^{-2}$ s$^{-1}$ at energies E $>$400 GeV
(i.e. about 3.5 times the Crab Nebula emission in the same energy range) 
seems to indicate that the
variability is due to the hardening/softening of the electron
spectrum, and not to the increase/decrease of the electron
density. Their model predicts for the second flare 
the Inverse Compton hump slightly shifted towards higher
energies and a VHE flux
a factor $\geq$2 larger with respect to the first one. 

Unfortunately
there were no VHE data included in their multiwavelength analysis
after June 8 because
the moonlight hampered the Cherenkov telescopes measurements. 
The VHE observation was actually made for such a very important 
flaring episode by the ARGO-YBJ experiment.

The ARGO-YBJ experiment, located at the Yangbajing Cosmic Ray
Laboratory (Tibet, P.R. China, 4300 m a.s.l., 30$^{\circ}$ 06'
38'' N, 90$^{\circ}$ 31' 50'' E), since December 2007
is performing a continuous monitoring of the sky in the
declination band from -10$^{\circ}$ to 70$^{\circ}$.

In this paper we present our observation of Mrk421 in flaring state
during 2008. After a summary of the data collected during the most
active phase (February-June),
we focus our discussion on the results obtained for the June flares 
in the framework of the Donnarumma et al. (2009) findings.

\section {The ARGO-YBJ experiment}

The ARGO-YBJ detector is constituted by a central carpet
$\sim$74$\times$ 78 m$^2$, made of a single layer of Resistive
Plate Chambers (RPCs) with $\sim$92$\%$ of active area, sorrounded
by a partially instrumented ($\sim$20$\%$) area up to
$\sim$100$\times$110 m$^2$. The apparatus has a modular structure,
the basic data acquisition element being a cluster (5.7$\times$7.6
m$^2$), made of 12 RPCs (2.8$\times$1.25 m$^2$). Each
chamber is read by 80 strips of 6.75$\times$61.8 cm$^2$ (the
spatial pixels), logically organized in 10 independent pads of
55.6$\times$61.8 cm$^2$ which are individually acquired and
represent the time pixels of the detector. The full detector is
made of 153 clusters for a total active surface of $\sim$6600
m$^2$ \citep{Aie06}. 

ARGO-YBJ operates in two independent
acquisition modes: the {\em shower mode} and the {\em scaler mode}
\citep{Aie08}. In the following we refer to the data recorded
in shower mode. 
A simple, yet powerful, electronic logic has been implemented
to build an inclusive trigger \cite{Alo04}.
This logic is based on a time correlation between the pad signals
depending on their relative distance. In this way, all the shower events giving
a number of fired pads 
N$_{pad}\ge$ N$_{trig}$ in the central carpet in a time window 
of 420 ns generate the trigger.
This trigger can work with high efficiency down to N$_{trig}$=20,
keeping negligible the rate of random coincidences.

The time of each
fired pad in a window of 2 $\mu$sec around the trigger time
and its location are recorded
and used to reconstruct the position of the shower
core and the arrival direction of the primary particle \cite{DiS07}.
In order to perform the time calibration of the 18360 pads, a
software method has been developed \cite{Aie09a}.

The detector is in stable data taking
with the trigger condition N$_{trig}$=20 and a duty cycle $\geq
85\%$. The trigger rate is $\sim$3.6 kHz with a dead time of 4$\%$.

\section{Detector performance}

The angular resolution and the pointing accuracy of the detector
have been evaluated by using the Moon shadow, i.e. the deficit of
cosmic rays in the Moon direction. The shape of the shadow
provides a measurement of the detector Point Spread Function
(PSF), and its position allows the individuation of possible pointing biases.
ARGO-YBJ observes the Moon shadow with a sensitivity of about 10
standard deviations per month for events with a multiplicity
N$_{pad} \geq$40 and zenith angle $\theta <$50$^{\circ}$,
corresponding to a proton median energy E$_p\sim$1.8 TeV \cite{DiS08}.

According to the Moon shadow data, the PSF of the detector is Gaussian for
N$_{pad} \ge$100, while for lower multiplicities it can be
described with an additional Gaussian, which contributes for about 20\%. 
When the PSF is a Gaussian with r.m.s. $\sigma$, the opening
angle  $\psi$ containing $\sim$71.5$\%$ of the events
maximizes the signal to background ratio for a
point source with a uniform background, and is equal to 1.58 $\sigma$.

The semi-aperture $\psi$ is found to be 
2.59$^{\circ} \pm$ 0.16$^{\circ}$,  1.30$^{\circ} \pm$ 0.14$^{\circ}$
and  1.04$^{\circ} \pm$ 0.12$^{\circ}$ 
for N$_{pad}\geq$40, 100 and 300, respectively,
in agreement with expectations from Monte Carlo simulations.

This measured angular resolution refers to cosmic ray-induced air
showers. The angular resolution for $\gamma$-induced events has been
evaluated by simulations and results 
smaller by $\sim$30-40$\%$, depending on  N$_{pad}$, due
to the better defined time profile of the showers. 

The relation between the observed pad/strip multiplicity spectrum
and the primary energy spectrum has been studied with cosmic ray
showers,
by means of a full Monte Carlo simulation, including the CORSIKA code 
\cite{Hec98} to describe
the shower development in the atmosphere, and a code based
on the GEANT package \cite{Gea93} to simulate the detector response.
Primary particles have been sampled from the energy spectra of
H\"orandel (2003).
The measured strip multiplicity spectrum is in good agreement
with the one predicted by the simulation \cite{Aie09b}.

\section{Data Analysis}

The dataset for the analysis of Mrk421 presented in this paper
contains all showers with N$_{pad} \geq$40 and zenith angle less
than 40$^{\circ}$. No event selection based on the shower core position 
and no gamma-hadron discrimination have been applied in this work.

A sky map in celestial coordinates (right ascension and declination) with
0.1$^{\circ}\times$0.1$^{\circ}$ bin size, centered on the source
location, is filled with the detected events. In order to extract
the excess of $\gamma$ rays coming from the source, the cosmic ray
background must be carefully estimated and subtracted from the
event map. 

The background is evaluated with the {\em
time swapping} method \citep{Ale92}. For each detected
event, N "fake" events are generated by replacing the original
arrival time with new ones, randomly selected from a buffer that
spans a time T of data taking. 
We chose T $\sim$ 3 hours to minimize the
systematic effects due to the environmental parameters variations.
Changing the time, the fake events maintain the same declination of 
the original event,
but have a different right ascension. With these events a new sky
map (background map) is built. 
The number of fake events generated for each event is
N = T(hr)$\times$15$\times$cos($\delta$)/2$\psi$,
where $\psi$ is the radius of the observational window in degrees (see 
below) and $\delta$ is the declination of the source.
In this way the average number of fake events falling in the 
observational window is $\sim$1.

The two maps are then "integrated" over a
circular area of radius $\psi$, i.e. every bin is filled with the
content of all bins whose center has an angular distance less than
$\psi$ from its center,
with $\psi$ = 1.7$^{\circ}$, 0.9$^{\circ}$ and  0.6$^{\circ}$
for N$_{pad}\geq$40, 100 and 300, respectively.

Finally the integrated background map is subtracted to the
corresponding integrated event map, obtaining the "source map",
where for every bin the statistical significance of the excess is
calculated.

With this procedure, however, 
since also the source events are used in the time swapping procedure,
the obtained background at the source position
is slightly overestimated, and the signal underestimated. 
This underestimation increases with the observational window size, ranging 
from $\sim$4 to 10$\%$ of the signal, depending on the N$_{pad}$ interval.
The observed event rate is then corrected using the appropriate factor.

The whole procedure has been tested with the Crab Nebula, the
standard candle for VHE astronomy. At the Yangbajing latitude the
Crab culminates at a zenith angle $\theta_{culm} = 8.1^{\circ}$ and
it is observable every day for 5.8 hours with a zenith angle
$\theta <$ 40$^{\circ}$.

The Crab Nebula was observed from 2007 December 13 to 2009 August
8, for a total of 3150 on-source hours, 
obtaining a signal with a statistical
significance of 7.6 standard deviations for N$_{pad}\geq$40.
The average number of gamma rays detected per day 
is 156.6$\pm$20.6 for N$_{pad}\geq$40. 

To evaluate the energy spectrum, we 
simulate a source in the sky following the daily path of the
Crab Nebula, and estimate the number of events expected 
in different N$_{pad}$ intervals, as a function of the spectrum parameters.

Assuming a power law spectrum in the 0.1-80 TeV
energy range, the best fit to the data is 
dN/dE = (4.1$\pm$0.6) $\times$ 10$^{-11}$ (E/1 TeV)$^{-2.7\pm 0.2}$ 
photons cm$^{-2}$s$^{-1}$ TeV$^{-1}$, in agreement with our previous
measurement \citep{Ver09} and with observations by other detectors, 
such as H.E.S.S. \citep{Aha06}, MAGIC \citep{Alb08} and Tibet AS-$\gamma$
\citep{Ame09}.
This result confirms the reliability of the simulation procedure and of the
energy calibration of the detector.

Concerning the energy range sampled in the Crab Nebula measurement,
about 84$\%$ of the detected events 
comes from primary photons of energies greater than 300 GeV, while only
8$\%$ comes from primaries above 10 TeV.

The same analysis was performed for Mrk421. This source
culminates at the ARGO-YBJ location at a zenith angle
$\theta_{culm} = 8.1^{\circ}$, and it is observable every day for
6.4 hours with a zenith angle $\theta <$ 40$^{\circ}$.

We evaluate the Mrk421 spectrum from day 41 to 180 of 2008, when the X-ray flux
showed the most intense flares.
In this period (754 observation hours)
the signal has a statistical significance of 5.8 standard deviations.

We assume a power law spectrum multiplied by an exponential factor  e$^{-\tau(E)}$
to take into account the absorption of gamma rays in 
the Extragalactic Background Light (EBL), 
with the values of the optical depth $\tau(E)$ given by Raue $\&$ 
Mazin
(2008).

The best fit spectrum obtained is:
dN/dE  = (3.0$\pm0.5$) $\times$ 10$^{-11}$(E/1.5 TeV)$^{-2.4\pm0.3}$ 
e$^{-\tau(E)}$ photons cm$^{-2}$ s$^{-1}$ TeV$^{-1}$.
The integral flux above 1 TeV is (3.6$\pm0.6$) 
$\times$ 10$^{-11}$ photons cm$^{-2}$ s$^{-1}$, 
almost twice the Crab Nebula one, i.e. 2.1 $\times$ 10$^{-11}$ 
photons cm$^{-2}$ s$^{-1}$, according to Aharonian et al. (2006).

The values of the spectral index and of the gamma ray flux averaged over this
140 days period, support the
correlation between spectral hardness and flux intensity
reported by Krennrich et al. (2002) and Albert et al. (2007), 
based on observations of Mrk421 in different activity states.

A complete account of the observations on Mrk421 will be reported in
a dedicated paper.

\section{The June 2008 flares }

As mentioned in the Introduction, two different flares have been observed
from Mrk421 in June 2008, the first one peaking in X-rays on June 4-6 and 
the second one on June 11-13. 
Concerning VHE gamma rays, Cherenkov telescopes 
data are available only for the first
flare. An energy spectrum for E $\geq$400 GeV has been provided by VERITAS 
for June 6.
 
Since the ARGO-YBJ sensitivity does not allow the observation of a flux 
a few times larger than the Crab one 
in only one day (i.e. during one transit of the source in the
detector field of view), we integrated the measurement over 3 days.

Fig.1 shows the rate of events with N$_{pad} \ge$100 observed
by ARGO-YBJ from June 3 to June 15, averaged over 3 days, compared with
the X-ray flux measured by RXTE/ASM\footnote[1]{
Public quick-look results 
(http://xte.mit.edu/asmlc/ASM.html)} 
in the 2-10  keV energy range. 
A correlation between the gamma ray and X-ray light curves is clearly visible.
During the days June 11-13, when the maximum of the second flare occurred, 
the excess of events from Mrk421 reached a statistical significance
of 3.8 standard deviations.

Beside the statistical error, this measurement could be affected by 
a systematic uncertainty due to the background evaluation, 
that is the most delicate step of the analysis.
In order to estimate this effect, a completely different
procedure for the background calculation has been implemented, using 
the so-called {\em equi-zenith angle} method \cite{Ame05}.
In this method the events collected in 10 off-source windows
of the same size of the on-source window, and aligned on both sides
of the same zenith angle belt, are used to obtain the
background. 
A detailed study of the two methods in the same sky region
has shown that on average they give
significances of the excesses consistent 
within 0.7 standard deviations, corresponding to about 20$\%$ uncertainty
on the flux estimate of the observed signal.

The event rate as a function of the minimum pad multiplicity,
obtained integrating the data during June 4-6 for the first flare 
(17.9 hours) and during June 11-13 for the second one (18.2 hours),
is shown in Fig.2.
On the same plot, the two solid lines represent the expected rates according 
to the Donnarumma et al. (2009) model, obtanined by a simulation
procedure. The SED proposed by this model has been corrected for the 
EBL absorption using 
the parameters given by Raue $\&$ Mazin (2008) in order to have 
the flux at Earth. Then, using the absorbed spectrum, 
we simulated a source moving along the Mrk421 
path on the sky, and evaluated
the number of events expected in the detector, for different N$_{pad}$
thresholds.
The complete simulation procedure (which includes 
the gamma ray showers propagation in the atmosphere, and 
the detector response)
has been tested evaluating the Crab Nebula flux, 
as shown in the previous section.

In the limit of the statistical accuracy of this measurement, our data suggest
for both flares a gamma ray flux 
higher than that expected by the model, 
indicating in particular a possible hardening of the spectrum
during the second flare.

Considering the first flare, the integral flux measured by ARGO-YBJ
above 1 TeV is about 1.5 times higher than
the model based on the VERITAS measurement,
but still marginally consistent with it.
The apparent disagreement between ARGO-YBJ and VERITAS can be 
likely attributed 
to the non-coincidence of the data taking periods of the two detectors
and to the well known small variability time scale of the source.
The VERITAS data refer to June 6, while the ARGO-YBJ data are
integrated over 3 days, from June 4 to 6.
Furthermore, given the difference in longitude of the two detectors
($\sim$160$^{\circ}$) and the fact that they observe the source
during few hours around the culmination time, 
they can never observe simultaneously the same object.

The disagreement of our data with the model is more significant for the
second flare.
In order to evaluate the spectral behaviour in this period,
we assume again a power law spectrum multiplied by the EBL absorption factor  
e$^{-\tau(E)}$.
From our fitting procedure we obtain:
dN/dE  = (3.2$\pm1.0$) $\times$ 10$^{-11}$(E/2.5 TeV)$^{-2.1^{+0.7}_{-0.5}}$ 
e$^{-\tau(E)}$ photons cm$^{-2}$ s$^{-1}$ TeV$^{-1}$.

This spectrum is shown in Fig.3 as a solid line. 
The shaded band in the figure represents the 1 $\sigma$ statistical error.
The systematic errors are mainly 
related to the background evaluation, as discussed previously, and to 
the uncertainty in the absolute energy scale. According to our estimate,
they globally affect the quoted fluxes for $\lesssim$ 30$\%$.

Due to the low statistics, our data cannot constrain the shape of the spectrum
above $\sim$8 TeV. 
Nevertheless the obtained flux appears, for energies $>$2 TeV, 
significantly larger than predicted by Donnarumma et al. (2009),
while the spectrum slope is consistent with that measured by 
the Whipple Cherenkov telescope during the 2000/2001 observing season
for a flare of comparable intensity ($\sim$ 7 times the Crab Nebula flux),
also shown in Fig.3 (dataset III, Krennrich et al. (2002)).

The integral flux measured above 1 TeV during June 11-13 
is $\sim$6 times larger than the Crab one, making this
flare one of the most powerful ever observed from Mrk421.

\section{Discussion and conclusions}

Mrk421 has been continuously monitored by ARGO-YBJ
since December 2007, showing an average VHE flux about twice the 
Crab Nebula one from
February to June 2008, and decreasing afterwards.

Two strong flares in June 2008 have been
observed in a multiwavelength campaign from optical to TeV energies.
ARGO-YBJ measured the spectra of Mrk421 above 0.3 TeV during the second flare, 
completing the multifrequency observations.
A clear correlation between the gamma ray intensity measured by ARGO-YBJ
and the X-ray flux measured by RXTE/ASM is found.

The ARGO-YBJ data, although averaged over 3 days, appear to support
in both episodes a gamma ray flux higher than that predicted
in the analysis of Donnarumma et al. (2009).
However, considering the short time scale variability of Mrk421,
it has to be noticed 
that our observation time is not fully coincident
with the period referred to by the theoretical curves (June 6 for the 
first flare and June 12-13 for the second one).

The intensity of the second flare allows us to assess its spectral
shape. 
The deabsorbed spectrum can be fitted by a power law 
$\propto$ E$^{-2.1^{+0.7}_{-0.5}}$ extending up to several TeV.
This spectrum appears definitively harder than that
predicted on the basis of June 12-13 data collected up to GeV energies. 

On the contrary, our data follow the behaviour of the
energy spectra measured during different activity states
by the Whipple Cherenkov telescope.
In particular, the ARGO-YBJ data fully satisfy the relation between
the spectral index and the flux obtained analyzing the 
measurements of Mrk421 since 1995 (see Fig.3 of Krennrich 
et al. (2002)).
This result indicates that this correlation is a long term property
of the source, as previously suggested by the Whipple collaboration.
 
A global analysis of all the data collected 
during the 2008 June 11-13 flare, including the present findings,
could be used to check the compatibility of the
observed phenomenology with current models for VHE photon emission
in the jets of AGNs.

\acknowledgments

We are grateful to the authors of Donnarumma et al. (2009), 
in particular to Marco Tavani and the AGILE team, 
for helpful discussions and for providing us
the Mrk421 broadband data relative to the period under study.

We also acknowledge the essential supports of 
W.Y. Chen, G. Yang, X.F. Yuan, C.Y. Zhao,
R.Assiro, B.Biondo, S.Bricola, F.Budano, A.Corvaglia, B.D'Aquino, R.Esposito,
A.Innocente, A.Mangano, E.Pastori, C.Pinto, E.Reali, F.Taurino and A.Zerbini, 
in the installation, debugging and maintenance of the detector.

%
%

\clearpage

\begin{figure}
    \includegraphics[height=0.4\textheight]{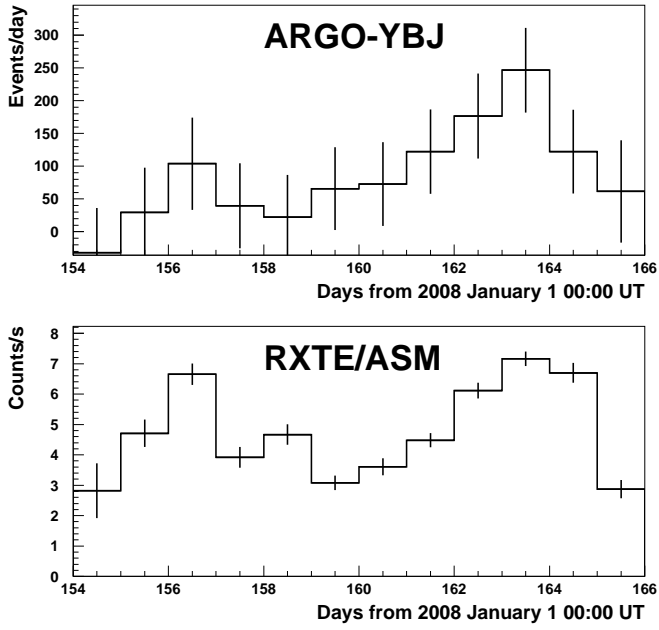}
\caption{Upper panel: rate of excess events with  N$_{pad} \ge$100 
observed from Mrk421 by ARGO-YBJ as a function of time
from 2008 June 3 00:00 UT to June 15 00:00 UT.
Each bin contains the rate averaged over the 3 days interval centered
on that bin.
Lower panel: daily counting rate of RXTE/ASM.}

\end{figure}


\begin{figure}
    \includegraphics[height=.4\textheight]{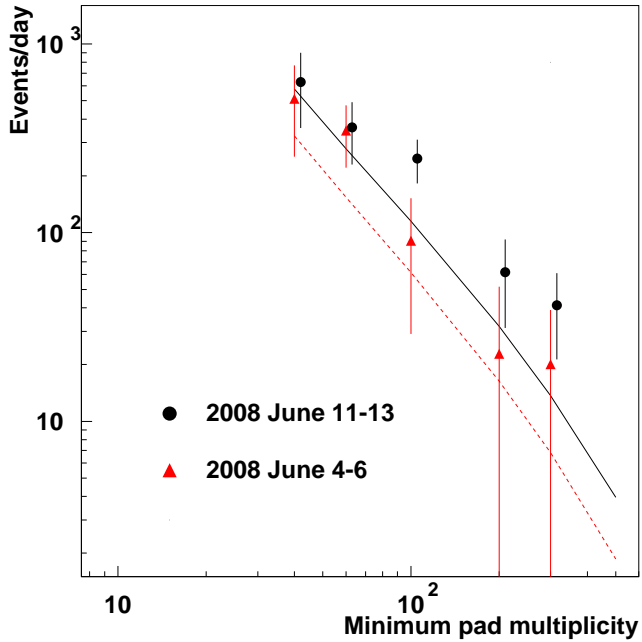}
\caption{Rates of the excess events observed from Mrk421 by ARGO-YBJ 
as a function of the event minimum
pad multiplicity on 2008 June 4-6 and June 11-13 (triangles and 
circles, respectively).
Expected rates according to the Donnarumma et al. (2009) model for June 
6th and June 12-13 (dashed and solid lines, respectively).}

\end{figure}

\begin{figure}
    \includegraphics[height=.4\textheight]{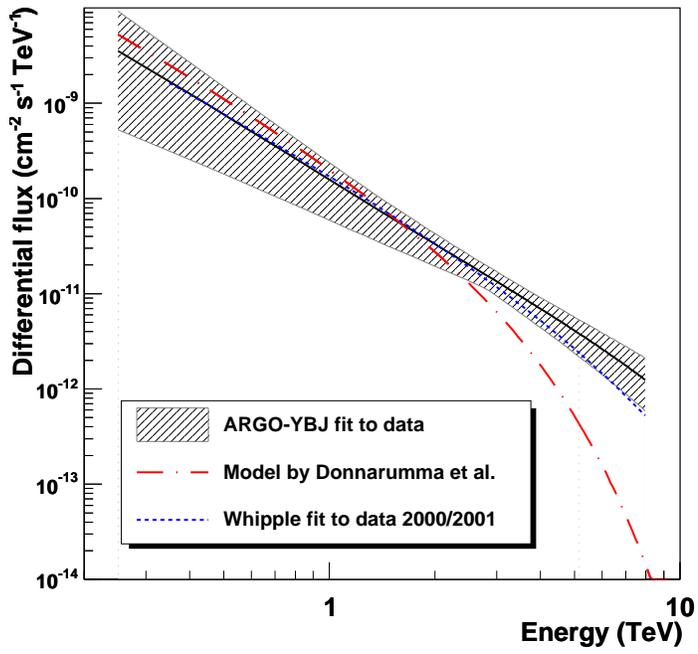}
\caption{
Gamma ray flux from Mrk421 measured by ARGO-YBJ on 2008 June 11-13
(solid line). The shaded band represents 1 standard deviation error.
The dot-dashed line shows the flux according to the
model by Donnarumma et al.(2009) for the second flare (June 12-13).
The dotted line shows the spectrum measured by Whipple \citep{Kre02}
during a previous flare of similar intensity (see text for details).}
\end{figure}


\begin{thebibliography}{99}

\bibitem[Aharonian et al. 2006]{Aha06} 
Aharonian, F., et al. 2006, A$\&$A, 457, 899
\bibitem[Aielli et al. 2006]{Aie06} 
Aielli, G. et al. 2006, NIM A, 562, 92
\bibitem[Aielli et al. 2008]{Aie08} 
Aielli, G., et al. 2008, Astrop. Phys., 30, 85
\bibitem[Aielli et al. 2009a]{Aie09a} 
Aielli, G., et al. 2009a, Astrop. Phys., 30, 287
\bibitem[Aielli et al. 2009b]{Aie09b} 
Aielli, G., et al. 2009b, Phys. Rev. D, 80, 092004
\bibitem[Albert et al. 2007] {Alb07} 
Albert, J., et al. 2007, \apj, 663, 125
\bibitem[Albert et al. 2008]{Alb08} 
Albert, J., et al. 2008, \apj, 674, 1037
\bibitem[Alexandreas et al. 1993] {Ale92} 
Alexandreas, D.E., et al. 1993, NIM A, 328, 570
\bibitem[Aloisio et al. 2004]{Alo04} 
Aloisio, A., et al. 2004, IEEE Transaction on Nuclear Science, 51, 1835
\bibitem[Amenomori et al. 2005]{Ame05} 
Amenomori, M., et al. 2005, \apj, 633, 1005
\bibitem[Amenomori et al. 2009]{Ame09} 
Amenomori, M., et al. 2009, \apj, 692, 61
\bibitem[Dermer et al. 1992]{Der92} 
Dermer, C.D., Schlickeiser, R., \& Mastichiadis, A. 1992, A\&A, 256, L27
\bibitem[Di Sciascio et al. 2007]{DiS07} 
Di Sciascio, G., et al. 2007, Proc. 30th ICRC, Merida, Mexico 
(arXiv:0710.1945)
\bibitem[Di Sciascio et al. 2008]{DiS08} 
Di Sciascio, G., et al. 2008, Proc. Vulcano Workshop 2008, Vulcano, 
Italy (arXiv:0811.0997)
\bibitem[Donnarumma et al. 2009]{Don09} 
Donnarumma, I., et al. 2009, \apj, 691, L13
\bibitem[Fossati et al. 1998]{Fos98} 
Fossati, G., Maraschi, L., Celotti, A. et al. 1998, MNRAS, 299, 433
\bibitem[Fossati et al. 2008]{Fos08} 
Fossati, G., Buckley, J. H., Bond, I. H., et al. 2008, \apj, 677, 906
\bibitem[GEANT 1993]{Gea93} 
GEANT - Detector Description and Simulation Tool  1993, CERN Program 
Library, W5013
\bibitem[Ghisellini et al. 1998]{Ghi98} 
Ghisellini, G., Celotti, A., Fossati, G., et al. 1998, MNRAS, 301, 451
\bibitem[Heck et al. 1998]{Hec98} 
Heck, D., Knapp, J., Capdevielle, J.N., Shatz, G., \& Thouw, T. 1998, Forschungszentrum Karlsruhe Report No. FZKA 6019
\bibitem[H\"orandel 2003]{Hor03}
H\"orandel, J.R. 2003, Astropart. Phys., 19, 193
\bibitem[Krennrich et al. 2002]{Kre02}
Krennrich, F., et al. 2002, \apj, 575, L9
\bibitem[M\"ucke et al. 2003]{Muc03}
M\"ucke, A., Protheroe, R.J., Engel, R., Rachen, J.P., \& Stanev, T. 2003, Astroparticle Physics, 18, 593
\bibitem[Padovani \& Giommi 1995]{Pad95}
Padovani, P., \& Giommi, P. 1995, \apj, 444, 567
\bibitem[Punch et al. 1992]{Pun92} 
Punch, M., Akerlof, C. W., Cawley, M. F., et al. 1992, Nature, 358, 477
\bibitem[Raue \& Mazin 2008]{Rau08}
Raue, M., \& Mazin, D. 2008, Int. J. Mod. Phys. D, 17, 1515
\bibitem[Sambruna et al. 1996]{Sam96} 
Sambruna, R. M.,Maraschi, L., \& Urry, C. 1996, \apj, 463, 444
\bibitem[Vernetto et al. 2009]{Ver09}
Vernetto, S. et al. 2009, Proc. of 2nd Roma International Conference on 
Astro-Particle Physics, in press
\bibitem[Wagner 2008]{Wag08} 
Wagner, R.M. 2008, PoS (BLAZARS2008), 63, 013 (arXiv:0809.2843)
\end{thebibliography}
\end{document}